\documentclass[aps, prb,showpacs,twocolumn, floatfix]{revtex4}
\usepackage{amsmath}
\usepackage{amssymb}
\usepackage{mathtools}
\usepackage{epsfig}
\usepackage{graphics}
\usepackage{color}
\usepackage{bm}
\usepackage{amssymb}
\usepackage{graphicx}
\begin{document}

\title{Nodeless Versus Nodal Scenarios of Possible Triplet Superconductivity
in the Quasi-One-Dimensional Layered Conductor
Li$_{0.9}$Mo$_6$O$_{17}$}

\author{O. Sepper}
\author{A.G. Lebed$^*$}

\affiliation{Department \ of \ Physics, \ University \ of \
Arizona, \ 1118 \ E. \ 4th \ Street, \ Tucson, \ AZ   \ 85721, \
USA}

\begin{abstract}
We consider the problem of the orbital upper critical magnetic
field, parallel to the most conducting axis of a
quasi-one-dimensional layered superconductor. It is shown that
superconductivity can be destroyed through orbital effects at
fields much higher than the so-called Clogston-Chandrasekhar
paramagnetic limiting field, $H_p$, provided that superconducting
pairing of electrons are of a triplet nature. We demonstrate that
the superconducting state of the quasi-one-dimensional layered
conductor, $\mathrm{Li_{0.9}Mo_6O_{17}}$, is well described by the
suggested theory. To this end, we consider two competing
scenarios: 1: a superconducting order parameter without zeros on
the Fermi surface, and 2: one with zeros on the Fermi surface  -
both are shown to lead to destruction of superconductivity at a
magnetic field, $H^x_{c_2}$, five times higher than $H_p$. With
recent experimental measurements on the
$\mathrm{Li_{0.9}Mo_6O_{17}}$ favoring the nodeless order
parameter, we present a strong argument supporting triplet pairing
in this compound.
\end{abstract}


\maketitle

\section{Introduction}

Detailed studies of the upper critical magnetic fields that lead
to the destruction of superconductivity in type-II superconductors
can provide essential information as to the nature of
superconductivity in a given compound. Theoretical analysis of the
upper critical magnetic fields along different directions in
superconducting crystals is crucial in highly anisotropic,
quasi-one and quasi-two dimensional (Q1D and Q2D) materials, where
such fields depend on the orientation, and can reveal fundamental
properties, such as the pairing symmetry of the superconducting
state. In a magnetic field, destruction of superconductivity that
results from the breaking of the Cooper pairs can manifest itself
through two distinct mechanisms, as the field couples to both
electron's charge and spin. In the first case, the magnetic field
alters the orbital wavefunctions of electrons, leading to the
orbital pair-breaking effect. Superconductors can experience this
Meissner effect irrespective of their pairing nature. The second
mechanism that leads to destruction of superconductivity in a
magnetic field is due to Pauli spin-splitting, as pairing of
electrons in spin-singlet states (spins anti-aligned, total spin
$s=0$) becomes energetically unfavorable. In this case, the
difference in Pauli energy levels, $\Delta E=2\mu_BH_p$, is of the
order of the superconducting energy gap, $\Delta$, where $\mu_B$
is the Bohr magneton, and $H_p$ is the so-called
Clogston-Chandrasekhar paramagnetic limiting field.\cite{Clogston}
A detailed analysis shows that
$H_p=\frac{\Delta}{2\sqrt{2}\mu_B}\approx 1.83 \ T_c \
(\mathrm{Tesla}/K^o)$, based on the BCS result\cite{BCS} of
$\Delta=3.53 \ k_BT_c$. In the case of the recently
examined\cite{Mercure} Q1D compound, Li$_{0.9}$Mo${_6}$O$_{17}$,
with $T_c = 2.2K$, this result gives $H_p\approx 4T$. We also note
that a paramagnetic limiting field of $H_p\approx 3.1T$ in this
compound has been extracted experimentally\cite{Mercure} from the
Pauli  susceptibility and the specific heat jump at $T_c$. This
value is five times smaller than the measured upper critical
field, $H^x_{c2} \approx 15T$, parallel to the most conducting
axis.

Survival of superconductivity for the upper critical magnetic fields greatly exceeding
$H_p$, suggests the possibility of spin-triplet pairing (where the total spin of a Cooper
pair, $s=1$) - a rather rare and intriguing phenomenon in unconventional superconductivity.
In contrast to singlet pairing, Cooper pairs in triplet superconductors can be insensitive to
Pauli splitting, the orbital pair-breaking effect being the prevalent mechanism that destroys
superconductivity.  In this regard, highly anisotropic, Q1D layered conductors have attracted
considerable attention from theorists and experimentalists alike. Important candidates for
unconventional superconductivity include the organic superconductors\cite{Jerome, Bechgaard}
that have been experimentally investigated since 1980. Initial experiments performed on the Q1D
superconductors $\mathrm{(TMTSF)_2X}$ ($\mathrm{X=PF_6}$ and $\mathrm{ClO_4}$), called
the Bechgaard salts, alluded to their unconventional nature.\cite{Lebed-1, Choi, Bouf, Takigawa}
Interest in these compounds was further intensified due to the possible existence of such peculiar
phenomena as reentrant superconductivity,\cite{Lebed-2, Lebed-3, DMS, Lee-Naughton} as well as
the Larkin-Ovchinnikov-Fulde-Ferrel (LOFF) phase,\cite{Larkin-Ovch, Fulde-Ferrel, Lebed-4, Buzdin}
and hidden reentrant superconductivity.\cite{Lebed-5}.  Currently, the leading candidate for triplet
superconductivity is the heavy-fermion compound\cite{UPt3} $\mathrm{UPt_3}$, while strong
evidence in favor of triplet pairing\cite{Sigrist-1, Maeno} has been found in $\mathrm{Sr_2RuO_4}$,
as theoretical studies of the latter have been motivated by similarities to triplet pairing in superfluid
$\mathrm{^3He}$. As for the members from the family $\mathrm{(TMTSF)_2X}$, NMR measurements
of the Knight shift provide evidence for $d$-wave like pairing\cite{ClO4, Yonezawa-1} for $\mathrm{X=ClO_4}$,
while the pairing nature of $\mathrm{X=PF_6}$, although initially hypothesized to be spin-triplet, has
not yet been unequivocally settled.\cite{PF6, Lee-Naughton} Investigation of the superconducting
state in Q1D conductors poses considerable general interest as more new compounds with unconventional
pairing symmetry (the possibility of triplet pairing, other exotic phases) are being experimentally investigated.

In our paper, we study two scenarios of triplet electron pairing in the Q1D layered superconductor
Li$_{0.9}$Mo${_6}$O$_{17}$. Using Gor'kov's equations for unconventional superconductivity,
\cite{Mineev, Sigrist-Ueda, Lebed-Yamaji} we obtain the so-called gap equations for superconducting
order parameters with and without zeros on the Q1D Fermi surface. We show quite generally that in
the absence of paramagnetic limiting the orbital pair-breaking effects lead to destruction of superconductivity
in a Q1D layered conductor at fields much higher than the Clogston-Chandrasekhar limit, $H^x_{c2} \gg H_p$,
with ${\bf H}$ aligned along the most conducting ($\hat{x}$) crystallographic axis, provided that the inter-plane
distance is less than the corresponding coherence length, $\xi_z$. This is in contrast to the common belief\cite{Mercure}
stipulating that the orbital destructive effects are minimized for fields parallel to the most conducting axis, and thus
are not able to destroy the superconducting phase. We define the band and superconducting parameters of
Li$_{0.9}$Mo${_6}$O$_{17}$ and show that, indeed, the coherence length of Cooper pairs perpendicular to the
planes, $\xi_z$, is greater than the inter-plane separation.  Thus, the conducting layers in Li$_{0.9}$Mo${_6}$O$_{17}$ are well coupled, leading to a 3D anisotropic description of superconductivity. We compare our results with the experimental data, \cite{Mercure} and demonstrate that the Q1D superconductor Li$_{0.9}$Mo${_6}$O$_{17}$ is better described by the nodeless triplet order parameter, in contrast to the nodal case. The temperature dependence of the uppercritical field, $H^x_{c2}(T)$, obtained in this paper is in excellent quantitative and qualitative agreement with the
experiment.  Note that the case  of nodeless triplet superconductivity was considered before in brief in our Rapid
Communication.\cite{Lebed-Sepper}

This paper is organized as follows: In Section II, the Hamiltonian of electrons with Q1D anisotropic energy
spectrum in a magnetic field is introduced and the wave functions are calculated. In Section III, the Green's
functions of the Q1D electrons in a magnetic field are obtained and the rest of the section is devoted to the
general formalism of obtaining the so-called gap equation for a triplet superconducting order parameter in a
magnetic field. Section IV is devoted to the analysis and analytical simplification of the integral gap equations
and their numerical solutions for the nodeless and nodal triplet superconducting order parameters. In Section V,
the upper critical magnetic field as a function of temperature, $H_{c_2}(T)$, is extracted for both kinds of order
parameters, and compared to the recently measured experimental values for the Q1D conductor Li$_{0.9}$Mo${_6}$O$_{17}$,
concluding with subsequent arguments in favor of  triplet pairing described by the nodeless case. In Section VI,
we summarize and discuss the obtained results.

\section{Wavefunctions of Electrons with Q1D
Anisotropic Electron Spectrum in a Magnetic Field}

We begin by considering the tight binding model for a Q1D electron
spectrum of a layered conductor:

\begin{equation}
E(\textbf{p})=-2t_x\cos(p_xa_x)-2t_y\cos(p_ya_y)-2t_z\cos(p_za_z),
\end{equation}

\noindent where we set $\hbar \equiv 1$. In the equation above,
$a_i$ are the lattice constants, and $t_i$ are the transfer
integrals for electron wavefunctions along the crystallographic
axes. In the highly anisotropic Q1D layered conductor under
consideration, $t_x>>t_y>>t_z$, a fact that will allow us to
linearize the dispersion relation. Our initial step is to define
the appropriate Hamiltonian, and solve a Schr\"{o}dinger-like
equation to obtain the exact wavefunctions. To this end,
we consider a magnetic field parallel to the conducting chains
(along $\hat{\textbf{x}}$) of a Q1D layered conductor,
$\textbf{H}=H\hat{\textbf{x}}$. The vector potential
corresponding to this magnetic field can be chosen to be
$\textbf{A}=Hy\,\hat{\textbf{z}}$. Consider a Q1D Fermi surface
(FS) - two open, slightly corrugated sheets centered at at
$p_x=\pm p_F$, extending along $\hat{\bf{p}}_z$ depicted in Fig.1.
On the surface of constraint (i.e. the FS) the following
linearized relation holds (the $\pm$ correspond to the left/right
sides of the FS):

\begin{figure}[t]
\centering
\includegraphics[width=65mm, height=75mm]{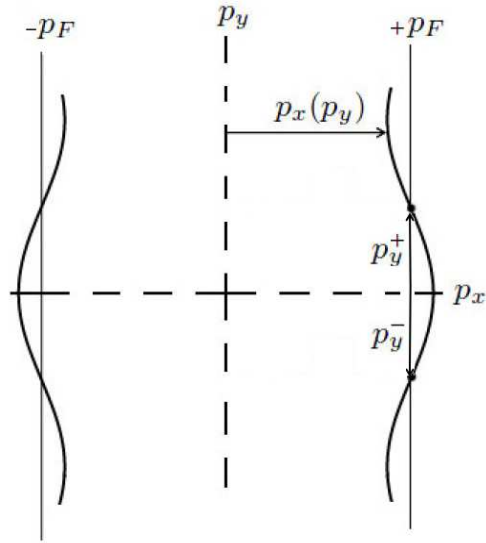}
\caption{Fermi surface for a Q1D layered conductor. Throughout the
text, the $\pm$ in expressions refer to the right (+) and left (-)
sheets of the Fermi surface.}
\end{figure}

\begin{equation}
p_x(p_y)=\pm p_F \pm \frac{2t_y}{v_F}\cos(p_ya_y),
\end{equation}

\noindent where  $v_F=2t_xa_x\sin(p_Fa_x)$ is the Fermi velocity
of electrons along the most conducting $\hat{\textbf{x}}$ axis.
Eq.(2)  implicitly defined $p_y$ as a function of $p_x$ on the Q1D
Fermi surface. Let $p_y^{\pm}$ stands for the two values (upper
and lower) of $p_y$ for which $p_x(p_y^{\pm})=p_F$. We further
define
\[v_y(p_x)=\partial \epsilon(\textbf{p})/\partial
p_y=2t_ya_y\sin[p_y(p_x)a_y]. \]

The energy dispersion relation can
be linearized near the left and right sheets of the FS. Measured
with respect to the Fermi energy, ${\epsilon=E-E_F}$, the
linearized dispersion relation takes the form

\begin{equation}
\epsilon^{\pm}(\textbf{p})=\pm v_y(p_y)
\left[p_y-p_y^{\pm}(p_x)\right] -2t_z\cos(p_za_z).
\end{equation}
To obtain the Hamiltonian in a magnetic field
$\textbf{H}=H\hat{\textbf{x}}$, the Pierels substitution method is
used:

\begin{equation}
p_y-p_y^{\pm}(p_x)\rightarrow -i\frac{\partial }{\partial y} \  \
\ \mathrm{and}  \  \  \  p_z \rightarrow p_z - \frac{e}{c}A_z ,
\end{equation}

\noindent where $A_z=Hy$ is the $z$-component of the vector
potential.

Using the substitution in Eq. (4) for the dispersion
relation in Eq. (3), as well as including the spin-dependent
interaction, the following Hamiltonian is obtained:

\begin{equation}
\hat{\epsilon}^{(\pm)}=\mp iv_y(p_y^{\pm})\frac{\partial}{\partial y} - 2t_z\cos\left(p_za_z- \frac{\omega_z}{v_F}y\right) \\
-2\mu_B s H,
\end{equation}

\noindent where $\omega_z=eHa_zv_F/c$, $\mu_B$ is the Bohr
magneton, and $s$ is the projection of the spin along the
direction of the magnetic field, $\hat{x}$. The simultaneous orbital
eigenfunctions of energy and momentum component $p_x$, with
eigenvalues $ \epsilon$ can be represented in the factored form

\begin{equation}
\Psi^{\pm}_{\epsilon, p_x}(x,y,p_z)=e^{\pm ip_xx}\,e^{\pm
ip_y^{\pm}(p_x)y} \, \psi^{\pm}_{\epsilon}(y,p_z).
\end{equation}
The wavefunctions, $\psi^{\pm}_{\epsilon}(y,p_z)$, are obtained from a
Schr\"{o}dinger-like equation $\hat{\epsilon}^{(\pm)} \, \psi^{\pm}_{\epsilon}(y,p_z) =\epsilon \, \psi^{\pm}_{\epsilon}(y,p_z)$:

\begin{multline}
\mp i v_y(p_y^{\pm})\frac{\partial
\psi^{\pm}_{\epsilon}(y,p_z)}{\partial y} =\left[ \epsilon +2t_z
\cos\left(p_za_z -\frac{\omega_z}{v_F}y\right)  + \right. \\
\biggl. +2\mu_B s H\biggr]\psi^{\pm}_{\epsilon}(y,p_z).
\end{multline}

The equation above admits exact solutions of the form

\begin{multline}
\psi^{\pm}_{\epsilon}(y,p_z)= \exp\left[\pm i
\frac{2t_z}{v_y(p_y^{\pm})}\int_0^y
\cos\left(p_za_z-\frac{\omega_z}{v_F}y'\right)\mathrm{d}y'\right] \\
\times \exp\left(\pm i \frac{\epsilon \, y}{v_y(p_y^{\pm})}\right)
\exp\left[\pm\frac{2 i \mu_B s H y}{v_y(p_y^\pm)}\right].
\end{multline}
Thus, the complete, normalized solutions for the wavefunctions are:

\begin{multline}
{\Psi^{\pm}_{ p_x}(\epsilon; x,y,p_z)= \frac{e^{\pm ip_x x}e^{\pm
i p_y^{\pm}y}}{\sqrt{2\pi |v_y(p_y^{\pm})|}} \exp\left[\pm i
\frac{\epsilon \, y}{v_y(p_y^{\pm})}\right] } \\
 \times \exp\left[\pm i \frac{2t_z}{v_y(p_y^{\pm})}
 \int_0^y\cos{\left(p_za_z-\frac{\omega_z}{v_F}y'\right)}\mathrm{d}y'\right] \\
\times \exp\left[\pm \frac{2 i\mu_B s H y}{v_y(p_y^\pm)}\right].
\end{multline}

\section{Green's Functions in a Magnetic Field and Triplet
Superconducting Pairing}

Having obtained the wave functions, we can calculate the Green's
functions. From the standard expression for the finite temperature
Green's function, we have

\begin{equation}
G^{\pm}_{i\omega_n}(\textbf{r},\textbf{r}')=
\sum_{\epsilon}\frac{\Psi^*_{\epsilon}(\textbf{r}')
\Psi_{\epsilon}(\textbf{r})}{i \omega_n-\epsilon},
\end{equation}
where $\omega_n= 2 \pi T (n+1/2)$ are the so-called Matsubara
frequencies. For convenience, we define the phase entering one of
the exponential factors in Eq. (9) as

\begin{equation}
\phi^{\pm}(y,p_z)=\frac{2t_z}{v_y(p_y^{\pm})}\int_0^y
\cos\left(p_za_z-\frac{\omega_z}{v_F}y'\right)\, \mathrm{d}y'.
\end{equation}
Substituting the wave functions from Eq. (9) into Eq.(10), and
converting the summation into integration over the energy
variable, we obtain the following expressions for the Green's functions:

\begin{multline}
{G^{\pm}_{i\omega_n}(x,x';y,y';p_z)=e^{\pm i p_x(x-x')}  e^{\pm i p_y^{\pm}(p_x)(y-y')}}  \\
\times \exp\left[\pm \frac{2i \mu_B s H (y-y')}{v_y(p_y^\pm)}\right] \hat{g}_{i\omega_n}^{\pm}(y,y';p_z),
\end{multline}

\noindent where the factor

\begin{equation}
\hat{g}_{i\omega_n}^{\pm}(y,y';p_z) = e^{ \pm i
\left[\phi^{\pm}(y,p_z)-\phi^{\pm}(y',p_z)\right]}
g_{i\omega_n}^{\pm}(y,y'),
\end{equation}

\noindent and the factor

\begin{equation}
g_{i\omega_n}^{\pm}(y,y')= \frac{1}{2\pi v_y(p_y^{\pm})} \int_{-\infty}^{\infty}
\frac{\exp\left[\pm i \frac{\epsilon (y-y')}{v_y(p_y^{\pm})}\right]}{i\omega_n-\epsilon}\mathrm{d}\,
\epsilon.
\end{equation}

In order to evaluate the integral in Eq. (14), a closed contour in
upper (lower) complex plane is used when $\omega_n>0$
$(\omega_n<0)$, which results in the following expressions for
$g_{i\omega_n}^{\pm}(y,y')$:

\begin{displaymath}
   g^+_{i\omega_n}(y,y') = \left\{
     \begin{array}{lr}
       \frac{-i \cdot\mathrm{sgn}(\omega_n)}{v_y(p_y^+)}
       \exp\left[\frac{-\omega_n(y-y')}{v_y(p_y^+)}\right] & : \omega_n(y-y')>0 \\
       0 & : \omega_n(y-y')<0
     \end{array}
   \right.
\end{displaymath}

\begin{displaymath}
   g^-_{i\omega_n}(y,y') = \left\{
     \begin{array}{lr}
       \frac{-i \cdot\mathrm{sgn}(\omega_n)}{v_y(p_y^-)}
       \exp\left[\frac{\omega_n(y-y')}{v_y(p_y^-)}\right] & : \omega_n(y-y')<0 \\
       0 & : \omega_n(y-y')>0
     \end{array}
   \right.
\end{displaymath}

\noindent Therefore, the expression for the Green's functions becomes, explicitly:

\begin{multline}
{G^{\pm}_{i\omega_n}(x,x';y,y';p_z)
=\frac{-i \cdot\mathrm{sgn}(\omega_n)}{v_y(p_y^{\pm})}} \\
\times e^{\pm ip_x(x-x')} e^{\pm ip_y^{\pm}(p_x)(y-y')}  \\
\times \exp\left[\pm i \frac{2t_z}{v_y(p_y^{\pm})}\int_{y'}^y\cos\left(p_za_z-\frac{\omega_z}{v_F}y''\right)\, \mathrm{d}y'' \right] \\
\times \exp\left[\frac{\mp \omega_n(y-y')}{v_y(p_y^{\pm})}\right] \exp\left[\pm \frac{2i\mu_B s H (y-y')}{v_y(p_y^\pm)}\right].
\end{multline}

\noindent In the above expression, the $+ \  (-)$ signs in the spin factor
$e^{\pm 2 i \mu_B s H (y-y')/v_y(p_y^+)}$ correspond to
electron in the up ($\uparrow$) or down ($\downarrow$) state,
respectively.

The derivation of the general expression for the superconducting
order parameter in the case of triplet pairing for a Q1D layered
conductor involves the use of Gor'kov's equation for
unconventional superconductivity. To this end, let us consider the
general expression \cite{Mineev} for a multi-component
superconducting order parameter:

\begin{multline}
{\Delta_{\alpha\beta}(\mathbf{k},\textbf{q})=-T\sum_n\sum_{k'k''q'}
V_{\beta\alpha,\lambda\mu}(\textbf{k},\textbf{k}')}\Delta_{\lambda\mu}(\mathbf{k}'',\mathbf{q}') \\
\times G_{\lambda}  \left( \mathbf{k}'+\frac{\mathbf{q}}{2}, \mathbf{k}''+\frac{\mathbf{q}}{2};
\omega_n\right) \\
\times G_{\mu}\left(-\mathbf{k}'+\frac{\mathbf{q}}{2}, -\textbf{k}''+\frac{\mathbf{q}}{2};
-\omega_n\right).
\end{multline}

\noindent In the above expression, $G(\mathbf{k}',\mathbf{k}'')$ are the
Fourier transformed Green's functions, the Greek subscripts
represent spin indexes, (that can take on two values represented
by $\uparrow$ or $\downarrow$) and a summation with respect to the
repeated index is implied. The spin dependent interaction,
$V_{\beta\alpha,\lambda\mu}(\textbf{k},\textbf{k}')$ can be
factorized in the absence of spin-orbit coupling as
$V_{\beta\alpha,\lambda\mu}(\textbf{k},\textbf{k}')
=V(\textbf{k},\textbf{k}')\Gamma_{\alpha\beta,\lambda\mu}$. In the
case of triplet superconducting pairing, the factors above have
the following properties:
$V(\textbf{k},\textbf{k}')=-V(-\textbf{k},\textbf{k}')=-V(\textbf{k},-\textbf{k}')$,
i.e. it is antisymmetric, whereas the factors
$\Gamma_{\alpha\beta,\lambda\mu}$ are symmetric under cyclic
interchange of the spin indexes, with non-zero values in the case of triplet
pairing being
$\Gamma_{\uparrow\uparrow,\uparrow\uparrow}=\Gamma_{\downarrow\downarrow,\downarrow\downarrow}=1$. 

We consider a triplet superconducting order parameter, $\Delta_t(\mathbf{k},\mathbf{q})$,
that is a linear combination

\begin{equation}
\Delta_{t}(\mathbf{k},\mathbf{q})=\Delta_{\uparrow\uparrow}(\mathbf{k},\mathbf{q})
+\Delta_{\downarrow\downarrow}(\mathbf{k},\mathbf{q}).
\end{equation}

\noindent Using the form of the interaction above and performing the
summation over the spin indexes in Eq. (16), we obtain the
following expression for
$\Delta_{\uparrow\uparrow}$ and $\Delta_{\downarrow\downarrow}$:

\begin{multline}
{\Delta_{\uparrow\uparrow}(\mathbf{k},\mathbf{q})= - T \sum_n  \sum_{k'k''q'}
\frac{ V(\mathbf{k},\mathbf{k}')}{2} } \times \left[ \right. \\
\left. \Delta_{\uparrow\uparrow}(\mathbf{k}'',\mathbf{q}')
\Omega_{\uparrow\uparrow}(\mathbf{k}',\mathbf{k}'',\mathbf{q})+
\Delta_{\downarrow\downarrow}(\mathbf{k}'',\mathbf{q}')
\Omega_{\downarrow\downarrow}(\mathbf{k}',\mathbf{k}'',\mathbf{q})\right],
\end{multline}

\noindent  where

\begin{multline}
\Omega_{\alpha\beta}(\mathbf{k}',\mathbf{k}'',\mathbf{q})
=\delta_{\alpha\beta}\, G_{\alpha}\left(\mathbf{k}'+\frac{\mathbf{q}}{2},\mathbf{k}''+\frac{\mathbf{q}}{2};
\omega_n\right) \\
\times G_{\beta}\left(-\mathbf{k}'+\frac{\mathbf{q}}{2},-\mathbf{k}''+\frac{\mathbf{q}}{2};
-\omega_n\right).
\end{multline}

\noindent In the expression above, $\alpha,\beta = \, \uparrow
\mathrm{or} \downarrow$. An expression similar to Eq. (18) is
obtained for
$\Delta_{\downarrow\downarrow}(\mathbf{k},\mathbf{q})$. Adding the
two quantities in Eq. (17), we obtain the general gap equation for
the triplet superconducting order parameter:

\begin{multline}
\Delta_t(\mathbf{k},\mathbf{q})=- T \sum_n  \sum_{k'k''q'} V(\mathbf{k},\mathbf{k}')
\Delta_t(\mathbf{k}'',\mathbf{q}') \\
\times  \left[
\Omega_{\uparrow\uparrow}(\mathbf{k}',\mathbf{k}'',\mathbf{q})+
\Omega_{\downarrow\downarrow}(\mathbf{k}',\mathbf{k}'',\mathbf{q})  \right].
\end{multline}

\section{The Triplet Superconducting Order Parameter:
Nodeless Versus Nodal Cases}

We consider two scenarios of triplet pairing in which
superconductivity is insensitive to Pauli paramagnetic effects.
The simplest such triplet superconducting order parameter takes
the form

\begin{equation}
\hat{\Delta}(p_x,y) = \hat{I}\mathrm{sgn}(p_x)\Delta(y),
\end{equation}

\noindent where $\hat{I}$ is a unit matrix in spin space, and the
function $\mathrm{sgn}(p_x)=\pm 1$ changes the sign of the order
parameter on the two sheets of the Q1D FS. The gap equation for
$\Delta(y)$ that determines  the upper critical field,
$H^x_{c2}(T)$, at which superconductivity is destroyed is obtained
by means of the general Eq.(20). We will first consider the case
when the order parameter does \emph{not} have zeros on the Q1D
Fermi surface [i.e., the order parameter (21)]. It is possible to
show that Eq.(20) for such a nodeless order parameter that
includes orbital destructive effects is reduced to the following
integral equation:

\begin{multline}
\Delta(y)=g \left\langle  \int_{|y-y'|>\frac{|v_y(p_y)|}{\Omega}}
\frac{  2\pi T \mathrm{d}y'}    {   v_y(p_y)\sinh\left[\frac{2\pi T |y-y'|}{v_y(p_y)}\right]} \Delta(y') \right. \\
\times \left. J_0\left\{\frac{8t_zv_F}{\omega_z v_y(p_y)}\sin\left[\frac{\omega_z(y-y')}{2v_F}\right]
\sin\left[\frac{\omega_z(y+y')}{2v_F}\right]\right\}\right\rangle_{p_y},
\end{multline}

\noindent where $\langle \cdot\cdot\cdot \rangle_{p_y}$ indicates
averaging over momentum $p_y$, introduced when the magnetic field
is parallel to the conducting axis.  Here, $g$ is a dimensionless
electron coupling constant, $\Omega$ is the cutoff energy, and
$\omega_z=eHa_zv_F/c$.

Eq. (22) is very general. Its solution defines a triplet
superconducting order parameter that includes the possibility of
re-entrant superconductivity\cite{Lebed-2} (in layered Q1D and
Q2D compounds) at very high magnetic fields and/or very low
temperatures, where the quantum nature of electron motion in a
magnetic field becomes important. Below, in analyzing the integral
in Eq. (22), we will work in the regime of relatively high
temperatures and relatively low magnetic fields, defined
respectively by the following conditions:

\begin{equation}
T\ge T^*(H)\approx \frac{\omega_z(H)v_y^0}{2\pi^2v_F},
\end{equation}

\begin{equation}
\omega_z(H)<<\frac{8t_zv_F}{v_y^0},
\end{equation}

\noindent where $v_y^0=2t_ya_y$. This is equivalent to neglecting
quantum effects resulting from Bragg reflection of electrons
moving along open FS, and amounts to replacing the first sine in
the above expression with its argument. It will be demonstrated
that the conditions in Eqs.(23) and (24) are well satisfied for
Li$_{0.9}$Mo$_6$O$_{17}$. These conditions render following
simplification to the arguments of the Bessel functions appearing
inside the integrals:

\begin{multline}
\Delta(y)=g\left\langle \int_{|y-y'|>\frac{|v_y(p_y)|}{\Omega}} \frac{2\pi T \mathrm{d}y'}{v_y(p_y)
\sinh\left[\frac{2\pi T |y-y'|}{v_y(p_y)}\right]} \Delta(y') \right. \\
\times \left. J_0\left\{\frac{4t_z(y-y')}{v_y(p_y)}\sin\left[\frac{\omega_z(y+y')}{2v_F}\right]\right\}
\right\rangle_{p_y},
\end{multline}

\noindent where $v_y(p_y)=v_y^0\sin(p_ya_y)$. This equation for
$\Delta(y)$ incorporates the description provided by the so-called
Lawrence-Doniah (LD) model,\cite{LD-Bulaevskii, LD-Klemm} where the coherence length perpendicular
to the conducting plane satisfies $\xi_z < a_z/\sqrt{2}$.
However, in the compound $\mathrm{Li_{0.9}Mo_6O_{17}}$,
the coherence length $\xi_z>a_z$, as will be shown below. Therefore, the LD
model does not apply, and the description of 3D anisotropic
superconductivity results. This fact further simplifies Eq. (25):

\begin{multline}
\Delta(y)=g\left\langle \int_{|y-y'|>\frac{|v_y(p_y)|}{\Omega}}  \frac{2\pi T \mathrm{d}y'}{v_y(p_y)
\sinh\left[\frac{2\pi T |y-y'|}{v_y(p_y)}\right]} \Delta(y') \right. \\
\times  \left. J_0  \left\{  \frac{2t_z\omega_z(y^2-{y'}^2)}{v_y(p_y)v_F} \right\}
\right\rangle_{p_y}.
\end{multline}

\noindent In order to recast Eq. (26) into a form appropriate for numerical analysis,
we employ the following change of variables:
\[ y' - y = \frac{v_y(p_y)}{v_y^0} z , \ \ \ \ \tilde{\omega}_z=\frac{v_y^0}{v_F}\omega_z, \]
with $\frac{v_y(p_y)}{v_y^0} = \sin\alpha$, after which, Eq. (26) can be expressed as

\begin{multline}
\Delta(y)=g\left\langle \int_d^\infty \frac{2\pi T \mathrm{d}z}{\sqrt{2t_z\tilde{\omega}_z}
\sinh\left[\frac{2\pi T}{\sqrt{2t_z\tilde{\omega}_z}}z\right]}\Delta (y + z\sin\alpha) \right. \\
\times \left. J_0  \left[ z ( 2y + z\sin\alpha)\right] \right\rangle_{\alpha},
\end{multline}

\noindent where cutoff distance $d = {\sqrt{2t_z\tilde{\omega}_z}/\Omega}$, and the averaging is now over the angular variable, $0<\alpha<2\pi$.
\ \newline

In the case where the triplet superconducting order parameter has
zeros on the FS, we take
\[\hat{\Delta}(\alpha, y)= \hat{I}  \mathrm{sgn}(p_x) \sqrt{2}\,  \sin(\alpha) \,  \Delta(y).\]

\noindent With analogous change of variables, the simplified integral equation
corresponding to the nodal case is:

\begin{multline}
\Delta(y) = g \left\langle  \int_{d}^{\infty}
\frac{2\pi T \mathrm{d}z}{\sqrt{2t_z\tilde{\omega}_z}
\sinh\left[\frac{2\pi T}{\sqrt{2t_z\tilde{\omega}_z}}z\right]} \, 2\sin^2\alpha \right. \\
\times \left. \Delta(y + z\sin\alpha) J_0  \left[ z ( 2y + z\sin\alpha)\right] \right\rangle_{\alpha}.
\end{multline}

\begin{figure}[t]
\centering
\includegraphics[width=90mm, height=60mm]{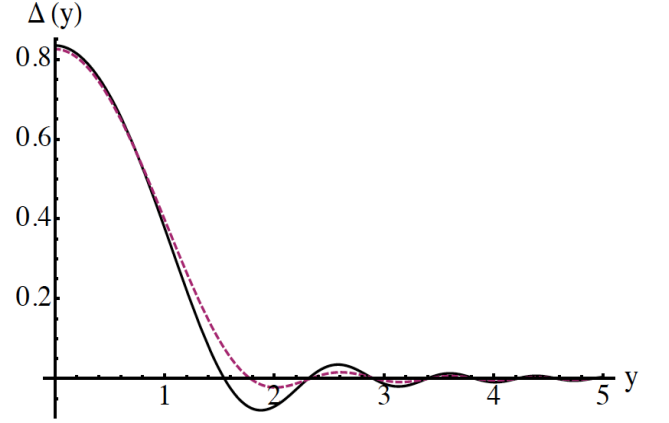}
\caption{Spatial dependences of the nodal (solid curve) and
nodeless (dashed curve) triplet superconducting order parameters
calculated at T=0.1 \ K.}
\end{figure}

\begin{figure}[t]
\centering
\includegraphics[width=90mm, height=60mm]{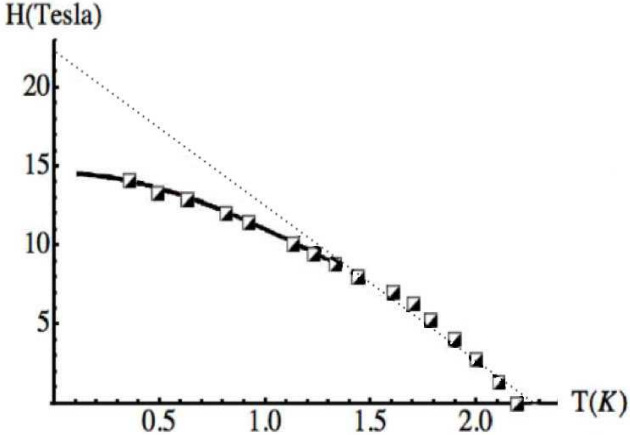}
\caption{Calculated temperature dependence of the upper critical
magnetic field, $H^x_{c2}(T)$, for the case of the nodeless order
parameter is represented by a solid line; squares represent
recently measured experimental values as reported in Ref. 3;
dashed line is the Ginzburg-Landau linear dependence valid for
$|T-T_c|<<T_c$.}
\end{figure}

\begin{figure}[t]
\centering
\includegraphics[width=80mm, height=65mm]{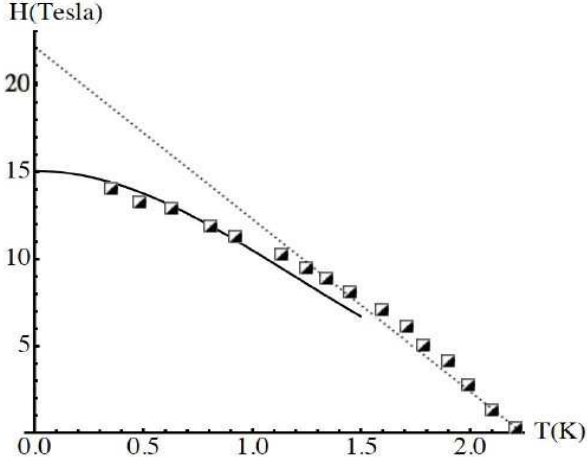}
\caption{Calculated temperature dependence of the upper critical
magnetic field, $H^x_{c2}(T)$, corresponding to the nodal order
parameter. Notations are the same as in Fig.3}
\end{figure}

\section{V. Calculated Upper Critical Magnetic Fields and their
Comparison with the Experiment}

The band and superconducting parameters for the Q1D electron spectrum of
$\mathrm{Li_{0.9}Mo_6O_{17}}$ can be determined from the GL slopes of the
measured upper critical fields, $H^i_{c2}(T)$, near $T_c$. The ratios of linearly
extrapolated zero-temperature upper critical fields along different axes are related to the
ratios of the corresponding coherence lengths through the Ginzburg-Landau relation,

\[ \frac{H^i_{c2}}{H^j_{c2}} = \frac{\xi_i}{\xi_j}. \]

\noindent Supplemented with the GL expression for the upper critical field,

\[H^i_{c2} = \frac{\Phi_0}{2\pi \xi_j \xi_k},\]

\noindent where $\Phi_0 = hc/2e$ is the magnetic flux quantum, the coherence lengths,
$\xi_i$, can be determined. In the vicinity of the superconducting transition
temperature,  ${|T-T_c|/T_c<<1}$, the anisotropic 3D Ginzburg-Landau expressions
for the upper critical magnetic field along different axes can be derived for both
nodeless and nodal order parameters.

It is possible to show that for the \emph{nodeless} order
parameter, Eq. (27) gives

\begin{equation}
H^x_{c2}(T)=\frac{4\pi^2cT_c^2}{7\zeta(3)et_yt_za_ya_z}\left(\frac{T_c-T}{T_c}\right),
\end{equation}

\noindent while the GL upper critical fields for directions perpendicular
to the most conducting axis are given by \cite{GLslopes-Lebed}

\begin{eqnarray}
H^y_{c2}(T)=\frac{  4\sqrt{2}\pi^2cT_c^2}{  7\zeta(3)ev_Ft_za_z} \left(\frac{T_c-T}{T_c}\right), \\
H^z_{c2}(T)=\frac{  4\sqrt{2}\pi^2cT_c^2}{  7\zeta(3)ev_Ft_ya_y} \left(\frac{T_c-T}{T_c}\right),
\end{eqnarray}

\noindent where $\zeta(3)$ is the value of the Riemann zeta function.
The GL coherence lengths for the nodeless case are:

\begin{equation*}
\xi_x=\frac{\sqrt{7\zeta(3)}v_F}{4\pi T_c}, \ \ \ \
\xi_y=\frac{\sqrt{7\zeta(3)}t_ya_y}{2\sqrt{2}\pi T_c},  \ \ \ \
\xi_z=\frac{\sqrt{7\zeta(3)}t_za_z}{2\sqrt{2}\pi T_c}.
\end{equation*}

For the \emph{nodal} order parameter, Eq. (28), the expression for
$H_{c2}^y$ remains identical to the nodeless case, while the
expressions for $H_{c2}^x$ and $H_{c2}^z$, are changed according
to:

\begin{eqnarray}
H^x_{c2}(T)=\frac{4\sqrt{2}\pi^2cT_c^2}{7\zeta(3)\sqrt{3}et_yt_za_ya_z}\left(\frac{T_c-T}{T_c}\right), \\
H^z_{c2}(T)=\frac{  8\pi^2cT_c^2}{  7\zeta(3)\sqrt{3}ev_Ft_ya_y} \left(\frac{T_c-T}{T_c}\right),
\end{eqnarray}

\noindent with the change only in the $\xi_y$ coherence length:

\[\xi_y = \frac{\sqrt{3\cdot 7 \zeta(3)}t_ya_y}{4\pi T_c}.\]
The band and superconducting parameters for the nodeless
case are summarized in Table I.

\begin{table}
\caption{Electron spectrum and superconducting parameters calculated in the case of nodeless order parameter.}
\begin{ruledtabular}
\begin{tabular}{c c  c c c}
$\bf{Li_{0.9}Mo_6O_{17}}$ & $\bf{\hat{x}}$ & $\bf{\hat{y}}$ & $\bf{\hat{z}}$  \\ \hline
$a_i(\AA)$ & 5.53 & 12.73 & 9.51  \\
$\xi_i(\AA)$ & 426 & 77& 20 \\
$t_i(K)$ & 370 & 41 & 14 \\
$v_i(cm/s)\cdot 10^6 $ & $v_F=5.3$ & 1.4 & 0.25 \\
\end{tabular}
\end{ruledtabular}
\end{table}

A numerical solution of Eqs. (27) and (28)  is implemented by
iteration, using a method of successive approximations both near
$T=0$, and for arbitrary (but small enough) values of $T$. The
solution for $\Delta(y)$ at $T=0.1 K$ for both nodeless and nodal
cases is shown in Fig.2. Note the qualitative difference compared
with the isotropic 3D superconductor: $\Delta(y)$ exhibits
decaying oscillations as a function of $y$. The period of these
oscillations is of the order of the coherence length $\xi_x$.
Furthermore, the temperature dependence of $\Delta$ can be shown
to be quadratic for small $T$. The solid lines in Fig.3 and Fig.4
correspond to the numerical solution to $H^x_{c2}(T)$ for nodeless
and nodal cases, respectively. The experimental data for
$\mathrm{Li_{0.9}Mo_6O_{17}}$ taken from Ref. 3 is overplotted as
squares, while the Ginzburg-Landau linear dependence (valid near
$T_c$) is plotted as a dashed line.

We can check the validity of approximations made in arriving at
the integral in Eq. (26) by using the values from the table above,
and the conditions in Eqs. (23) and (24). The results are:

\[T\ge T^* \approx 0.06K \ \ \ \ \ \mathrm{and}  \ \ \ \ \ \ \ H<<300T\]
These conditions are well satisfied in experiments of Ref. 3.
Furthermore, as the coherence length $\xi_z\approx 20 \AA >
a_z/\sqrt{2} = 6.7\AA$, i.e. is it much greater than the
interlayer spacing, the layered are well coupled, and the so
called Lawrence-Doniah model does not apply in this context. Thus,
our problem is that of anisotropic 3D superconductivity. As shown
previously by us \cite{Lebed-Sepper} for a magnetic field parallel
to the most conducting axis of a Q1D layered conductor, orbital
effects are capable of destroying superconductivity in the absence
of paramagnetic effects. Although both nodeless and nodal triplet
order parameters  (Fig.2) reproduce qualitatively similar results
for the phase diagram of $H^x_{c2}(T)$, the nodeless case is in a
better quantitative agreement with the data from Ref. 3.

As an additional concluding remark, we discuss another possible explanation of a very high
upper critical magnetic field, parallel to conducting axis in a Q1D superconductor.
As shown in Ref. 35, in a pure 1D singlet superconductor Pauli spin-splitting effects in a magnetic field do not destroy superconductivity at $T=0$ in arbitrarily high magnetic fields due to formation of the LOFF phase$^{14,15}$ in a form of the soliton superstructure. Nevertheless, in Ref. 36, it was shown that in a real singlet Q1D superconductor with electron spectrum (1), there exists paramagnetically limiting magnetic field for the LOFF phase even in the case where the orbital effects against superconductivity are negligible. The field that paramagnetically limits singlet superconductivity in a Q1D superconductor is evaluated$^{36}$ as

\begin{equation}
H^{LOFF}_p = 0.6 \ \sqrt{t_x/t_y} \ H_p \ .
\end{equation}

Substituting the corresponding parameters for the $\mathrm{Li_{0.9}Mo_6O_{17}}$
superconductor (see Table I), we obtain $H^{LOFF}_p \simeq 6 \ T$.
As shown in Ref. 16, the orbital effects against superconductivity decrease
this paramagnetically limiting field. Therefore, we conclude that the possible appearance of the LOFF phase in the framework of a singlet scenario of superconductivity is very unlikely to be responsible for the very large experimental upper critical field, $H^x_{c2} \approx 15 \ T$, in the $\mathrm{Li_{0.9}Mo_6O_{17}}$ superconductor.

\section{Summary}

In this paper, we have explored two competing scenarios within the
framework of triplet superconducting pairing in a Q1D layered
conductor. We have demonstrated how in a parallel magnetic field,
superconductivity can be destroyed through orbital effects, and
have calculated $H^{x}c_2(T)$ when the triplet order parameters
has and does not have nodes on the FS. Our findings show that the
nodeless order parameter leads to the temperature dependence of
upper critical field that is in a better quantitative agreement
with the experimental data in Ref. 3. In particular, such a
nodeless triplet order parameter is consistent with the large
value of the experimentally observed specific heat jump at the
superconducting transition in zero field. Note that all our
calculations have been done within a validity of the so-called
Fermi-liquid picture. In this context, it is important that the
quadratic dependence of low temperature magnetoresistance is
reported in Ref. 3, which we consider as a main argument in favor
of the Fermi liquid description. In more details, we state that
the low temperature regime $(T<T_c\approx 2.2K)$ at which our
calculation are performed avoids the Luttinger liquid behavior
that is expected to emerge in Q1D conductors at higher
temperatures,\cite{Wang, Santos, Wakeham} as was observed in
Li$_{0.9}$Mo$_6$O$_{17}$.
\newline

One of us is thankful to N.N. Bagmet, N.E. Hussey, and S. Mazumdar
for useful discussions. This work was supported by the NSF under
Grant DMR-1104512.

$^*$Also at: L.D. Landau Institute for Theoretical Physics, 2
Kosygina Street, Moscow 117334, Russia.


\end{document}